\documentclass{article}

\begin{document}

\begin{center}
{\Large Solitons in a Three-Wave System with Intrinsic Linear Mixing and
Walkoff}

\bigskip

Arkady Kaplan$^{1}$ and Boris A. Malomed$^{2}$
\end{center}

\bigskip

$^{1}$CeLight Inc., 12200 Tech Road, Silver Spring, MD 20904, USA

$^{2}$Department of Interdisciplinary Studies, Faculty of Engineering, Tel
Aviv University, Tel Aviv 69978, Israel

\bigskip

\begin{center}
{\large Abstract}
\end{center}

\bigskip

A modification of the usual Type-II second-harmonic-generation model is
proposed, which includes two additional features: linear conversion and
walkoff (group-velocity difference) between two components of the
fundamental-frequency (FF) wave. Physical interpretations of the model are
possible in both temporal and spatial domains. In the absence of the
intrinsic walkoff, the linear mixing makes real soliton solutions stable or
unstable, depending on the relative sign of the two FF components. Unstable
solitons spontaneously re-arrange themselves into stable ones. Fundamental
solitons change their shape (in particular, they develop chirp) but remain
stable if the intrinsic walkoff is included. In addition, quasi-stable
double-humped solitary waves are found in the latter case.

\newpage

\section{Introduction and formulation of the model}

Spatial and temporal solitons in various models with quadratic ($\chi ^{(2)}$)
) nonlinearity, including the second-harmonic-generating (SHG) and more
general three-wave (3W) systems, have attracted a lot of attention, see
reviews \cite{Torner,Jena,Aussie} and a special volume \cite{Boardman}. The
simplest (degenerate) SHG model involves only two waves, the
fundamental-frequency (FF) one and its second harmonic (SH). In a general
situation, one is dealing with a 3W system, that involves two ``daughter''
waves and a pump; it is known that the 3W system also gives rise to soliton
solutions \cite{3W}. A case of great practical interest is a more special
version of the 3W system, known as the Type-II SHG model, in which both
``daughters'' represent two polarizations of the FF wave, while SH has a
single polarization. The latter model corresponds to the most typical
experimental conditions \cite{Torner}, and its three-component soliton
solutions (also called vectorial solitons) were studied in detail \cite
{vector,Peschel,Dawn,Jena}.

Various modifications of the Type-II system were introduced, and the soliton
solutions in them were studied [see, e.g., Refs. \cite{modified}]. These
modifications were focused on adding extra quadratic terms to the model's
equations. Another possibility which has not yet been considered, except for
Ref. \cite{Mak} (see below), is to introduce linear coupling (mixing)
between the two FF components. The simplest physical realization of this
possibility may be found in terms of the Type-II SHG process in a fiber-like
birefringent waveguide, which is subjected to a twist that mixes the two FF
polarizations. Note that twist-induced linear mixing between two linear
polarizations is a well-known feature of optical fibers with the Kerr ($\chi
^{(3)}$) nonlinearity \cite{twist}. Of course, application of the twist to a
monocrystalline waveguide in which SHG takes place is problematic, but an
effective twist, without applying any mechanical torque, can be created in
an evident way in the practically important case when SHG itself is induced
by means of periodic poling of the host medium (see a review \cite{Fejer}).

A system of equations to describe the Type-II SHG in a birefringent medium
in the presence of the linear mixing can be derived as a straightforward
generalization of the model developed in Ref. \cite{Peschel}: 
\begin{eqnarray}
i\delta A_{z}+ib_{1}A_{\tau }+\left( 1/2\right) A_{\tau \tau }-\beta
A+B^{\ast }C &=&\kappa B,  \label{A} \\
i\delta ^{-1}B_{z}+ib_{2}B_{\tau }+\left( 1/2\right) B_{\tau \tau }-\beta
^{-1}B+A^{\ast }C &=&\kappa A,  \label{B} \\
i\sigma C_{z}+\left( 1/2\right) DC_{\tau \tau }-\alpha C+AB &=&0,  \label{C}
\end{eqnarray}
where the asterisk and subscripts stand, respectively, for the complex
conjugation and partial derivatives, the evolution in the temporal domain is
implied, $\tau $ being the standard reduced-time variable \cite{Agrawal},
complex fields $A(z,\tau )$, $B(z,\tau )$ and $C(z,\tau )$ represent the two
FF components and SH wave, respectively, $\beta $ is a phase-birefringence
parameter, $\alpha $ measures the phase mismatch between the FF and SH
waves, $\delta >0$ is another parameter taking into regard asymmetry between
the two FF components \cite{Peschel}, $D$ is a relative SH/FF dispersion
coefficient, and $\sigma >0$ determines a relative propagation constant, all
these parameters being real.

New (in comparison with the known model) terms introduced in Eqs. (\ref{A})
and (\ref{B}) are the linear-mixing ones with the real coefficient $\kappa$,
and group-velocity-birefringence (temporal-walkoff) terms with real
coefficients $b_{1,2}$ [the reference frame is defined so that the
group-velocity term vanishes in Eq. (\ref{C})]. Note that in the absence of
the linear mixing, $\kappa =0$, the walkoff terms can be easily removed from
the model by means of phase transformations and change of the reference
frame, therefore the model elaborated in Ref. \cite{Peschel} did not include
these terms. However, these terms are irreduceable if $\kappa \neq 0$.

The model may also be interpreted in the spatial domain, with $\tau $ is
replaced by the transverse coordinate $x$, $b_{1,2}$ being spatial walkoff 
\cite{walkoff} parameters. In fact, the linear coupling between the FF
components in the 3W model in the spatial domain (a planar optical waveguide
with the $\chi ^{(2)}$ nonlinearity) may be induced in a simple way by means
of a Bragg grating (a system of parallel scores) written on the waveguide 
\cite{Mak}. In the latter case, all the fields have the same polarization,
the difference between the two FF components being in the direction of their
Poynting vectors.

An objective of this work is to find fundamental-soliton solutions of the
model and test their stability by means of precise numerical simulations.
Prior to that, we will consider the model's linear spectrum, which is
necessary in order to realize what type of solitons may exist in it.

For $\kappa \neq 0$ but $b_{1,2}=0$, stationary soliton solutions to Eqs. 
(\ref{A}), (\ref{B}), and (\ref{C}) can be easily found. A dynamical test
will show that the stationary solitons may be both stable and unstable in
this case, depending on the relative sign of the two FF components, which is
a difference from the results reported in Ref. \cite{Peschel} for the
Type-II SHG model without the linear mixing ($\kappa =0$). With the
introduction of the walkoff parameters ($b_{1,2}\neq 0$), numerical solution
of the stationary equations becomes difficult, therefore we rely upon direct
simulations in this case: propagation of an initial solitary-wave
configuration, which is taken as a stable soliton for the same values of
parameters but with $b_{1}=b_{2}=0$, generates a stable fundamental soliton,
featuring intrinsic chirp, after a transient process. Additionally, we also
find quasi-stable double-humped solitary waves in the latter case.

\section{The linear spectrum}

Solitons are represented by solutions to Eqs. (\ref{A}) - (\ref{C}) of the
form 
\begin{equation}
A(z,\tau )=e^{iKz}a(\tau ),\,B(z,\tau )=e^{iKz}b(\tau ),\,C(z,\tau
)=e^{2iKz}c(\tau ),  \label{soliton}
\end{equation}
where $K$ is a real propagation constant, and $a(\tau )$, $b(\tau )$, and 
$c(\tau )$ are even functions (complex ones, in the general case) that must
vanish at $\left| \tau \right| \rightarrow \infty $. As is commonly known,
(bright) soliton solutions may only exist at values of $K$ that do not
overlap with the continuous spectrum of the FF part of the system. The
spectrum is determined by the substitution of the continuous-wave solution, 
\begin{equation}
A(z,\tau )=A_{0}\exp \left( iKz-i\omega \tau \right) ,B(z,\tau )=B_{0}\exp
\left( iKz-i\omega \tau \right) ,\   \label{cw}
\end{equation}
into linearized versions of Eqs. (\ref{A}) and (\ref{B}).

As for the SH component, a similar condition for the existence of solitons
(the non-overlapping with the continuous spectrum) is not always necessary
in it, since situations are possible when the SH equation may \emph{not} be
linearized, its quadratic term being everywhere (at all values of $\tau $,
including $\left| \tau \right| \rightarrow \infty $) on the same order of
magnitude as the linear terms. Such a situation is possible as the quadratic
and linear terms in the SH equation are composed of different fields: the
former one contains only the FF components, while the linear terms are
expressed in terms of the SH field, see Eq. (\ref{C}). Eventually, this
leads to a possibility of the existence of the so-called embedded solitons,
for which the SH propagation constant $2K$ [see Eq. (\ref{soliton})] indeed
falls into the linear spectrum of the SH wave \cite{embedded,Alan}.

The combination of the linear coupling and group-velocity difference between
the two FF components in Eqs. (\ref{A}) and (\ref{B}) makes the present
system somewhat akin to $\chi ^{(2)}$ models in which effective dispersion
and/or diffraction are induced by a Bragg grating, while the intrinsic
dispersion (or diffraction) are neglected. Models of this type, which have a
finite gap in their linear spectrum and, accordingly, give rise to \textit{\
gap solitons} supported by the quadratic nonlinearity, have attracted
considerable attention, see Refs. \cite{gap,gap2,Thomas,Mak}. However, the
present model, despite its similarity to the gap-soliton ones, has the
linear spectrum of a different type, which contains no finite gap but,
instead, the usual semi-infinite gap extending to $K\rightarrow +\infty$.
Indeed, taking, in order to avoid ponderous formulas, Eqs. (\ref{A}) and 
(\ref{B}) with $\delta =1$, $\beta =1$, $b_{1}=-b_{2}\equiv b$, and
substituting the expressions (\ref{cw}) into the linearized equations, one
can easily obtain the following expression for the linear spectrum: 
\begin{equation}
K=-\left( 1+\omega ^{2}/2\right) \pm \sqrt{\left( b\omega \right)
^{2}+\kappa ^{2}}.  \label{spectrum}
\end{equation}
It is straightforward to see that the combination of two branches of this
spectrum does not yield any finite gap [that would exist, in the form 
$\left( K+1\right) ^{2}<\kappa ^{2}$, if the dispersion terms in Eqs. (\ref{A})
and (\ref{B}), and hence the term $\omega ^{2}/2$ in Eq. (\ref{spectrum}),
were omitted]. On the other hand, all sufficiently large positive values of 
$K$ do not belong to the linear spectrum, forming the above-mentioned
semi-infinite gap (in the case $\kappa ^{2}>b^{4}$, it simply takes the form 
$K>0$).

Thus, the present system combines, to a certain extent, previously studied
ordinary soliton models and those which give rise to gap solitons, which
makes it interesting to search for solitons in it.

\section{Solitons in the model without intrinsic walkoff}

\subsection{Unstable solitons}

We start by seeking for soliton solutions to Eqs. (\ref{A}) - (\ref{C}) in
the case $b_{1,2}=0$ (no intrinsic walkoff, while the linear mixing is
present, $\kappa \neq 0$). The stationary version ($\partial /\partial z=0$)
of the system of Eqs. (\ref{A}), (\ref{B}) and (\ref{C}) then becomes real,
and, accordingly, stationary solutions are looked for in a real form by
means of the shooting method. Note that, in the absence of the linear mixing
($\kappa =0$), solutions differing by a relative sign of the FF components, 
$A$ and $B$, are, obviously, equivalent. However, this is not the case if 
$\kappa \neq 0$ (which is similar to the situation in a $\chi ^{(2)}$ model
of a dual-core waveguide, where linear terms couple FF fields in two cores 
\cite{dual-core}). In particular, it is obvious that Eqs. (\ref{A}) -- (\ref
{C}) can be derived from a Hamiltonian, its term which accounts for the
linear mixing being 
\begin{equation}
H_{\mathrm{mix}}=\kappa \int_{-\infty }^{+\infty }\left( A^{\ast }B+AB^{\ast
}\right) d\tau \,.  \label{mixing}
\end{equation}
In the case of real solutions, and taking, by definition, $\kappa >0$, one
may expect that the solutions with $\mathrm{sgn}A=\mathrm{sgn}B$ should be
unstable, as they make the term (\ref{mixing}) of the Hamiltonian positive,
while solutions with $\mathrm{sgn}A=-\mathrm{sgn}B$ may be stable, as they
yield $H_{\mathrm{mix}}<0$ (a known principle states that a solution which
makes the value of the Hamiltonian larger is likely to be unstable \cite
{Berge'}).

The application of the shooting method to the stationary real equations (\ref
{A}) - (\ref{C}) with $b_{1,2}=0$ indeed yields stationary solitons in a
broad parametric region, provided that the relative dispersion coefficient
is positive, $D>0$. A typical example of a solution with $\mathrm{sgn}A= 
\mathrm{sgn}B$, generated by the shooting method, is displayed in Fig. 1.

Direct simulations of the dynamical stability of the thus found stationary
solitons were performed by means of the implicit Crank-Nicholson scheme. As
is known \cite{Yevick}, this scheme has advantages over the split-step
method for very long beam-propagation simulations. To control the accuracy
of the direct simulations, we made use of the fact that the underlying
system (\ref{A}) -- (\ref{C}) conserves the net energy of the three fields, 
\begin{equation}
E=\int_{-\infty }^{+\infty }\left( \delta \left| A\right| ^{2}+\delta
^{-1}\left| B\right| ^{2}+\sigma \left| C\right| ^{2}\right) d\tau \,.
\label{E}
\end{equation}
Results of the simulations complied with the conservation of the integral 
(\ref{E}) to a very high accuracy.

The application of the Crank-Nicholson scheme to the soliton shown in Fig. 1
has produced a picture which is displayed, in the contour-plot form, in Fig.
2. This result, as well as many other runs of simulations for other
stationary solitons, demonstrate that, in accordance with the general
argument presented above, all the real solitons with $\mathrm{sgn}A=\mathrm{
\ sgn}B$ are unstable. It is relevant to compare a characteristic
propagation length $z_{\mathrm{stab}}$ before the onset of the instability,
which is $\simeq 80$ in Fig. 2, and the soliton period for the same pulse
(which is defined as the propagation length necessary for the change of the
internal phase of the soliton by $\pi /2$ \cite{Agrawal}), that can be
estimated for the soliton shown in Fig. 1, in terms of its width $W$, as $z_{
\mathrm{sol} }\sim \pi W^{2}\simeq 15$. This means that the unstable soliton
can pass $\sim 5$ soliton periods as a quasi-stable object. In a typical
experimental situation for spatial solitons, their diffraction length is 
$\sim 1$ mm, while the size of a sample is a few cm \cite{Torner}, hence both
the unstable soliton itself and its instability may be experimentally
observed in the spatial domain.

In fact, Fig. 2 displays a typical scenario of the development of the
instability of solitons with $\mathrm{sgn}A=\mathrm{sgn}B$: after a
relatively long period of a very slow ``latent'' growth of the instability,
an abrupt explosion occurs, as a result of which the soliton sheds off some
amount of radiation in the SH component, and rearranges itself into a new,
completely stable, soliton (these stable solitons will be described below).
Figure 3 shows the quasi-stable-propagation distance $z_{\mathrm{stab}}$ of
the unstable solitons vs. the coupling constant $\kappa $, for several
different values of the relative SH dispersion coefficient $D$. The fact
that, irrespective of the value of $D$, the distance $z_{\mathrm{stab}}$
diverges as $\kappa \rightarrow 0$ is quite natural, as in this limit we get
back to the usual model without the linear mixing, where the real solitons
are stable irrespective of the relative sign of their two FF components \cite
{Peschel}.

If SH dispersion coefficient $D$ is very small, the character of the
instability development becomes qualitatively different from that
illustrated by Fig. 2: as is seen in Fig. 4, in this case only the SH
component of the soliton survives. Note that, in the case $D=0$, Eqs. (\ref
{A}), (\ref{B}), and (\ref{C}) have an obvious solution corresponding to the
eventual state observed in Fig. 4: $A=B=0$, $C(z,\tau )=\exp \left( -i\alpha
\sigma ^{-1}z\right) \cdot c(\tau )$, where $c(\tau )$ is an arbitrary
function.

\subsection{Stable solitons}

In the case $b_{1}=b_{2}=0$, another class of real stationary soliton
solutions can be found (by means of the shooting method too), with opposite
signs of the two FF components ($\mathrm{sgn}A=-\mathrm{sgn}B$). A typical
example of such a soliton is shown in Fig. 5. In accordance with the
qualitative arguments given above [based on the sign of the coupling term 
(\ref{mixing}) in the Hamiltonian], the solitons of this type are found to be
completely stable in direct simulations of their evolution. An example is
displayed in terms of the contour plots in Fig. 6 (note that the full
propagation distance presented in Fig. 6 is $\sim 100$ soliton periods; in
fact, the stability was seen, in much longer simulations, to persist
indefinitely).

An accurate analysis of the stable pulses, the formation of which was
observed as a result of the development of the instability of the stationary
solitons with $\mathrm{sgn}A=+\mathrm{sgn}B$ (see Fig. 2), shows that the
appearing stable pulses are identical to the stable solitons of the type
considered here, with $\mathrm{sgn}A=-\mathrm{sgn}B$. Thus, these solitons
are really robust, playing a role of \emph{attractors} in the evolution of
unstable pulses (the existence of effective attractors in a conservative
nonlinear-wave system is possible due to radiation losses).

\section{Solitons in the system with the intrinsic walkoff}

\subsection{Single-humped solitons}

It is quite interesting to understand how solitons are modified if the
group-velocity (walkoff) terms are restored in Eqs. (\ref{A}) and (\ref{B}).
Search for stationary soliton solutions of the full system of Eqs. (\ref{A})
- (\ref{C}), which include the walkoff terms, turns out to be much harder
than in the case $b_{1}=b_{2}=0$, as it appears quite difficult to secure
convergence of results produced by the shooting method. Therefore, we
adopted an approach based on direct simulations of the full equations in the
following fashion: stationary solutions corresponding to stable solitons
(with $\mathrm{sgn}A=-\mathrm{sgn}B$), which were found above for the case 
$b_{1}=b_{2}=0$ (for instance, the soliton shown in Fig. 5), were used as
initial conditions for simulations of the evolution equations (\ref{A}) -- 
(\ref{B}) with the same values of all the parameters but $b_{1}$ and $b_{2}$.
Note that the introduction of the walkoff terms must essentially rearrange
the input pulses, as they are real, while stationary solutions to Eqs. (\ref
{A}) -- (\ref{B}) with $b_{1,2}\neq 0$ cannot be real.

A typical example of the evolution of the thus chosen input pulse is shown,
by means of contour plots, in Fig. 7 [in this figure, we display the
evolution of local powers $\left| A(\tau )\right| ^{2}$, $\left| B(\tau
)\right| ^{2}$, and $\left| C(\tau )\right| ^{2}$]. The propagation distance
in Fig. 7 is extremely large ($\simeq 200$ soliton periods), which was taken
in order to make it sure that a final soliton, if any, takes a sufficiently
well-established form. The result of the evolution is shown in Fig. 8.

A conclusion suggested by this and many other runs of simulations is that
stable solitons with intrinsic chirp establish themselves in the presence of
the intrinsic walkoff, although radiation shed off from the soliton in the
course of its self-adjustment separates from it very slowly. The latter
peculiarity can be understood. Indeed, the radiation tail attached to the
soliton in Fig. 8 (upper panel) is all built of the SH field, which has zero
group velocity in the underlying equation (\ref{C}), hence it does not
readily separate from the zero-velocity soliton. It seems very plausible
(although detailed consideration of the issue is beyond the scope of this
work) that Eqs. (\ref{A}) -- (\ref{C}) may also generate moving
(``walking'') solitons, in which case the separation of the soliton from the
radiation ``garbage'' would probably be faster.

To further check that the (quasi-) soliton (called this way because of the
radiation tail attached to it), whose formation and structure are shown in
Figs. 7 and 8, has long since completed any essential evolution, in Fig. 9
we show the evolution (vs. $z$) of the net energy (or intensity, in the case
of beams in the spatial domain) of each component of the soliton, i.e., 
\begin{equation}
\int_{-\infty }^{+\infty }\left| A(z,\tau )\right| ^{2}d\tau
,\,\int_{-\infty }^{+\infty }\left| B(z,\tau )\right| ^{2}d\tau
,\,\int_{-\infty }^{+\infty }\left| C(z,\tau )\right| ^{2}d\tau  \label{net}
\end{equation}
($\int_{-\infty }^{+\infty }$ is realized as the integral over the whole
simulation domain), and Fig. 10 displays the evolution of the total energy
defined by Eq. (\ref{E}). A very small initial loss of the total energy (see
Fig. 10) is explained by leakage across borders of the integration domain in
the process of the initial rearrangement of the soliton. Note that intensive
energy exchange between the $A$ and $C$ fields (see Fig. 9) is limited to
approximately the same initial stage of the evolution at which the energy
loss takes place; then, any tangible evolution ceases, in terms of the
integral field characteristics (\ref{net}).

\subsection{Double-humped structures}

The approach described above produces stable \textit{fundamental} solitons,
i.e., single-humped ones. On the other hand, it is well known that $\chi
^{(2)}$ models readily give rise to higher-order solitons -- first of all,
double-humped ones \cite{unstabledoublehumped} -- which, however, are always
unstable in standard models, including the 3W Type-II model \cite{Jena}.
Search for \emph{stable} double-humped solitons in various systems is a
problem of considerable interest for physical applications, see, e.g., Refs. 
\cite{stabledouble}. In fact, the first examples of (numerically) stable
one-dimensional double-humped solitons were found in 3W models combining 
$\chi ^{(2)}$ nonlinearity and linear coupling (which was induced by the
Bragg grating) between two components of the FF field \cite{Mak,Thomas}.
Moreover, the model introduced in Ref. \cite{Mak}, that seems to be closest
to the one considered in the present work, gives rise also to vast families
of double- and multi-humped embedded solitons \cite{Alan}, although the
stability of those solutions was not studied in detail.

We made an attempt to search for double-humped solitary-wave structures in
the present model. In the absence of the linear mixing ($b_{1,2}=0$), they
have never been found, which seems quite natural in view of the
above-mentioned results obtained in allied models. However, structures of
that type can indeed be found at finite (actually, quite small) values of 
$\left| b_{1,2}\right| $, and they seem to be nearly stable, although they
are not completely stationary.

To this end, the stationary version of Eqs. (\ref{A}) - (\ref{C}), produced
by the substitution of the waveforms (\ref{soliton}), was first solved
numerically with high but finite accuracy, starting from the numerically
exact solution for $b_{1}=b_{2}=0$, such as the one shown in Fig. 5, and
gradually increasing the parameter $b_{1}=-b_{2}\equiv b$. As it was
mentioned above, in the presence of the walkoff terms with $b_{1,2}\neq 0$
straightforward application of the shooting technique does not provide for
convergence of soliton solutions to indefinitely high accuracy, this is why
the accuracy was finite, as mentioned above. It was observed that if other
parameters keep constant values (for instance those which are mentioned in
the caption to Fig. 11), the increase of $b$ makes the (finite-accuracy)
soliton broader, and \emph{splitting} of the soliton's crest into two takes
place at $b=0.00274$. A typical example of the appearing double-humped
structure is shown, for a slightly larger value $b=0.003$, in the upper
panel of Fig. 11.

The simulated evolution of this structure over a very long propagation
distance shows that this solitary wave is not a genuine steady-state
solution, but it is quite close to being one. It keeps a well-pronounced
double-humped shape over, at least, $15$ soliton periods. This implies that
the double-humped structure is robust enough to be observed in\ an
experiment.

Further increase of $b$ makes the double-humped pulses still less localized,
and, eventually, permanent leakage of one of the FF components from the
pulse starts. It is difficult to find a critical value of $b$ at which this
pulse ceases to exist as a solitary-wave solution, as an extended ``tail''
of the FF field, the appearance of which signals the onset of the leakage,
has a vanishingly small amplitude when it emerges.

\section{Conclusion}

We have proposed a modification of the usual three-wave
second-harmonic-generation model which incorporates two features that are
new to the usual model: linear mixing between two components of the
fundamental-frequency wave, and a group-velocity mismatch (walkoff) between
them. Although the new system is akin to gap-solitons models, its linear
spectrum contains no finite gaps. In the temporal domain, the model may be
interpreted as the one adding an (effective) twist of the fiber-like
waveguide to the birefringence, the latter feature being typical for the
Type-II $\chi ^{(2)}$ systems. In the spatial domain, the two FF components
differ, physically, not by their polarizations, but rather by the
orientation of their Poynting vectors in a planar waveguide, the linear
coupling being induced by the Bragg grating.

In the absence of the intrinsic walkoff, the linear mixing induces a
difference between real soliton solutions with the opposite relative signs
between the two FF components, so that they are stable for one sign, and
unstable for the other. The development of the instability leads to
rearrangement of unstable solitons into stable ones. Adding the
intrinsic-walkoff terms, we have found that the evolution leads to formation
of stable chirped fundamental solitons, and, additionally, quasi-stable
double-humped solitary waves were found.

\section*{Acknowledgement}

This work was supported in a part by a grant from the Research Authority of
the Tel Aviv University.

\newpage

\begin{center}
{\Large Figure Captions}
\end{center}

Fig. 1. An example of a stationary real soliton solution with $\mathrm{sgn}
A= \mathrm{sgn}B$. The parameters are $\delta =2$, $\beta =0.778$, $\kappa
=0.2$, $\sigma =3$, $D=1$, $\alpha =0.156$, $b_{1,2}=0$ (no walkoff between
the two fundamental-frequency wave components). In this and subsequent
figures, the argument $x$ attached to the horizontal axis replaces the
variable $\tau $ for a case when the model is interpreted in the spatial
domain, where $x$ is the transverse coordinate (see the text).

Fig. 2. Evolution of the soliton shown in Fig. 1.

Fig. 3. The distance necessary for the onset of the instability of the
soliton with $\mathrm{sgn}A=\mathrm{sgn}B$ (as detected in the
second-harmonic component) vs. the linear-mixing constant $\kappa $.

Fig. 4. Evolution of the soliton in the case $D=0.03$, other parameters
taking the same values as in Fig. 1.

Fig. 5. An example of a stationary real soliton with $\mathrm{sgn}A=-\mathrm{
sgn}B$ for the same values of parameters as in Fig. 1.

Fig. 6. Evolution of the soliton shown in Fig. 5.

Fig. 7. Evolution of the input pulse identical to the soliton shown in Fig.
5 at the same values of parameters as in Fig. 5, except for 
$b_{1}=-b_{2}=0.1 $ (cf. Fig. 6, which pertains to the case $b_{1,2}=0$).

Fig. 8. Panels (a) and (b) display, respectively, the distribution of local
powers, $\left| A(\tau )\right| ^{2}$, $\left| B(\tau )\right| ^{2}$, and 
$\left| C(\tau )\right| ^{2}$, and local chirps, $\left( \phi _{A}\right)
_{\tau \tau }$, $\left( \phi _{B}\right) _{\tau \tau }$, and $\left( \phi
_{C}\right) _{\tau \tau }$ ($\phi $ stands for the phase of field), of the
three waves in the (quasi)soliton generated by the evolution process
displayed in Fig. 7.

Fig. 9. The net energies of the three components of the soliton, defined as
per Eq. (\ref{net}), vs. the propagation distance $z$, for the same case as
in Fig. 7.

Fig. 10. The total energy (\ref{E}) of all the three fields vs. $z$, shown
for the same case as in Fig. 9. Note that the energy loss, due to some
leakage through the edges of the integration domain, is very small (see
numerical values on the vertical axis).

Fig. 11. An example of a quasi-stable double-humped solitary-wave structure
found for $\delta =2$, $\beta =0.778$, $\kappa =0.2$, $\sigma =4$, $D=1$, 
$\alpha =0.156$, and $b_{1}=-b_{2}=0.003$. The upper and lower panels show,
respectively, the initial configuration at $z=0$, obtained as a
finite-accuracy shooting solution of the stationary equations, and the final
configuration obtained at $z=4000$. As well as in Fig. 8, the distribution
of the field powers across the solitary wave is shown here.

\end{document}